# Evaluation of Control/User-Plane Denial-of-Service (DoS) Attack on O-RAN Fronthaul Interface

Ferlinda Feliana*, Ting-Wei Hung*, Binbin Chen†, and Ray-Guang Cheng*
* National Taiwan University of Science and Technology † Singapore University of Technology and Design

*Abstract*—The open fronthaul interface defined by O-RAN ALLIANCE aims to support the interoperability between multi-vendor open radio access network (O-RAN) radio units (O-RU) and O-RAN distributed units (O-DU). This paper introduces a new tool that could be used to evaluate Denial-of-Service (DoS) attacks against the open fronthaul interface. We launched an array of control/user planes (C/U-Planes) attacks with the tool under different traffic types and data rates, and we evaluated their impacts on the throughput and block error rate (BLER) of real-world O-RAN systems with commercial hardware.

## I. INTRODUCTION

O-RAN ALLIANCE was founded to develop an intelligent, open, virtualized, and fully interoperable radio access network (RAN). Several open interfaces were defined by the O-RAN ALLIANCE. Among them, the open fronthaul interface plays a critical role in enabling the interoperability between multi-vendor O-RAN radio units (O-RU) and O-RAN distributed units (O-DU). The open fronthaul interface consists of four distinct planes: Control (C-Plane), User (U-Plane), Synchronization (S-Plane), and Management (M-Plane). The O-DU uses C-Plane messages to instruct the O-RU for processing the U-Plane messages exchanged between O-DU and O-RU. The U-Plane messages are used to carry the in-phase and quadrature-phase data (IQ data) in the fronthaul. The S-Plane messages are used to keep accurate timing and synchronization among network components. The M-Plane is for management operations [2] [3].

The Security Working Group of the O-RAN ALLIANCE [4] identified several potential threat models [4] and addressed four specific threats related to the C/U-Plane. The first threat, T-FRHAUL-02, highlights the potential risk of an unauthorized device on the Ethernet L1 interface gaining access to C/U-Plane traffic on the open fronthaul interface. The second threat, T-CPLANE-01, highlights the possibility of an adversary injecting a false C-Plane downlink message, claiming to originate from the associated O-DU. This injection can block the processing of corresponding U-Plane packets in O-RU, causing a temporary denial of service (DoS). The third threat, T-CPLANE-02, addresses the injection of a false C-Plane uplink message. The fourth threat, T-UPLANE-01, disclosed a potential threat from an attacker launching a DoS attack on the U-Plane. The testing and integration focus group (TIFG) is responsible for defining the O-RAN end-to-end test specifications [5]. Among them, the C-Plane eCPRI DoS attack is one of the key test items and requires a test tool to conduct an attack against the MAC address of the O-DU C-Plane. The tool needs to send eCPRI real-time control data messages over Ethernet in 3 volumetric tiers: 10 Mbps, 100 Mbps, and 1 Gbps. The source address of those DoS packets should support spoofed MAC addresses of O-RU(s), random source MAC addresses, and broadcast MAC addresses [26].

Several commercial tools are available to conduct the C-Plane DoS attack test, including the products offered by Keysight [8], Viavi [9], and Xena Networks [10]. However, as they are proprietary tools, it is difficult to customize them to support new tests. Open-source tools such as Metasploit [11] or Yersinia [12] do not support the eCPRI-based open fronthaul interface. One may utilize the packet capture (PCAP) replay method supported by tools having the PCAP replay feature, such as Metasploit's pcap_replay or tcpreplay [13]. However, users need to have a PCAP file containing the C-plane packets. In addition, the common packet size of a C-Plane packet is 64 bytes. Hence, it's also difficult for such specialized replay tools to generate DoS traffic at the volumetric tier of 1 Gbps. The limitations of existing tools highlight the need for specialized tools and methods to effectively test and secure open fronthaul in O-RAN networks.

The C-plane DoS attack problem [5] was first investigated in [14]. The authors utilized the fronthaul library provided by O-RAN Software Community (OSC) [15] to implement a C-plane DoS attack tool and used the tool to attack an emulated E2E environment. They found that around 20% and 35% of U-Plane messages were lost at the DU side if the attack rate reached 5 and 10 Gbps, respectively. They also found that a significant amount of U-plane data packets were dropped at the RU side even if the attack rate was 10 Mbps when the attack tool chose the spoofed O-DU MAC address as its source address. Under such a situation, it may also result in incorrect forwarding at the fronthaul gateway due to addressing confusion. Different from [14], this paper presented a new C/U-Plane DoS attack tool and investigated the impact of the DoS attacks in a real end-to-end O-RAN environment. We consider both the C-Plane and U-Plane DoS attacks to provide a comprehensive understanding of the vulnerabilities and potential threats that can arise in the open fronthaul interface. We conducted various attacks in an end-to-end environment using our developed tool, incorporating different traffic types, rates, and source MAC addresses. The impact of these attacks was investigated in terms of throughput, block error rate (BLER), and the operating states of both O-DU and O-RU. The main contributions of this work include:

- We developed the first open-source C/U-plane DoS at-

tack tool that is fully compliant with O-RAN E2E test specification [5] and allows easy extension. The source code of our tool is available publicly at https://github.com/bmw-ece-ntust/cu-plane-dos-attacker.

- We conducted a detailed empirical evaluation on two commercial end-to-end O-RAN systems and showed that our tool can effectively reveal different levels of protection in the systems under test.

The rest of the paper is organized as follows. Section II defines the system and threat model considered in this paper. The development of the C/U-Plane attack tool is explained in Section III. Section IV shows the experimental results. Concluding remarks and future work are given in Section V.

## II. SYSTEM AND THREAT MODEL

We consider a connection between O-DU and O-RU(s) through a Fronthaul Gateway. The C/U-Plane DoS attacker can either control a new device connected to the Fronthaul Gateway and break through the authentication method or compromise a component that is already connected with the Fronthaul Gateway. The attack tool is compliant with the test case in [5] with all criteria on volumetric, source MAC addresses, and message type supported. In addition, we added two new message types (U-Plane downlink and U-Plane uplink), a new case where the source MAC address is the same as the destination MAC address, and a new target that is O-RU to observe possible loopholes that exist in the system. The scope of this research is to observe how much of an impact it has on the end-to-end system when the handling of certain message types and source MAC addresses fails. We hence explored the impact of potential loss of availability mentioned in the threat ID T-FRHAUL-02 (with the limit of U-Plane and C-Plane), the threat ID T-CPLANE-01, and the threat ID T-UPLANE-01. The threat ID T-CPLANE-02 on C-Plane uplink is currently out of the scope of this research.

## III. C/U-PLANE DOS ATTACK TOOL

In this system, we design a Linux-based C/U-Plane DoS attack tool to generate C/U-plane DoS packets to the O-DU. The purpose of the C/U-Plane DoS Attack Tool is to send C/U-plane packets toward the target using specific settings (including different source MAC addresses, different volumetric tiers, etc.). We designed the tool to be extensible so that, in addition to the experiment conducted in this research, it can provide different test cases (e.g., fuzzing) for future usage to test C/U-planes. Figure 1 shows the key building blocks of our C/U-plane attack tool. There are three main building blocks of the attack tool. The first part is the PCAP file generator, with the main function to generate the packets that will be transmitted or used to attack. The second part is the O-RAN-FH packet dissector built on Python as a custom Scapy dissector supporting dissection of O-RAN fronthaul C-plane packet Type-1 with extType-1 and U-plane packet. The third part is the transmission of packets from the previously generated PCAP file.

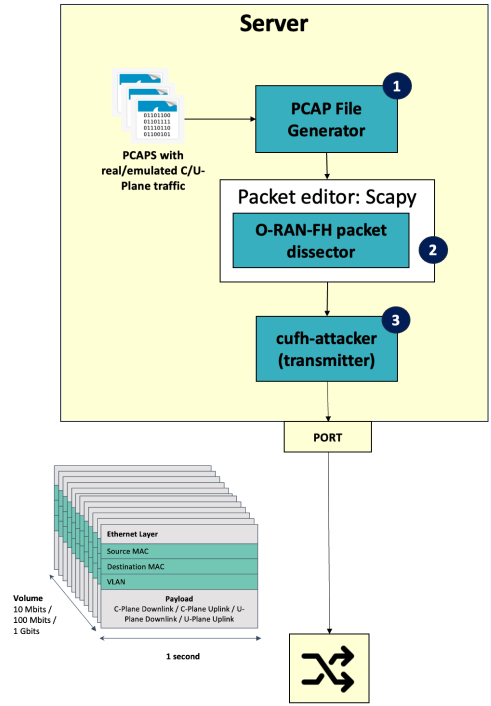

Fig. 1. Our C/U-Plane DoS attack tool architecture

The PCAP file generator is written in Python as a command line tool to generate the packet(s) to be used in the attack. The main function of this tool is to choose packet(s) from available PCAPs containing C/U-plane packets previously captured from real or simulated traffic and create a new PCAP from the chosen packet(s) with the amount of packet fulfilling the volume (in Mbits) user wants to generate. The optional function includes the editing of the VLAN tag, source MAC address, and destination MAC address. It also can set randomized source MAC addresses. The O-RAN-FH packet dissector is a custom layer module that can be added to Scapy [17]. Its purpose is to allow the editing of C/U-Plane packet fields. It can be used to edit certain fields of packets in the PCAP file generated by the PCAP file generator. The cufh-attacker is built based on DPDK-burst-replay [18] functions to burst PCAP dump on one or more NIC port(s) by using DPDK (Data Plane Development Kit). In this paper, DPDK Release 19.11 was used. The cufh-attacker module is built by modifying the DPDK-burst-replay code to fulfill the testing needs. The addition allows the cufh-attacker to send packets continuously every second, either at a constant volume or on an incremental basis (increasing the volume per input Mbps every second). Considering the need to directly change the source MAC address, destination MAC address, and VLAN tag without going through the generation of the PCAP file, functions to edit those three fields are supported. The relationship between PCAP files generated by the PCAP file generator and cufh-attacker is from the parsing of the user's input. Three types of messages, named C-Plane downlink, U-Plane downlink, and U-Plane uplink, were supported. Note that in our attacks, we purposely send C-Plane downlink traffic towards the O-DU and as we will show in our experiment that such traffic can still cause impact with certain spoofed MAC address.

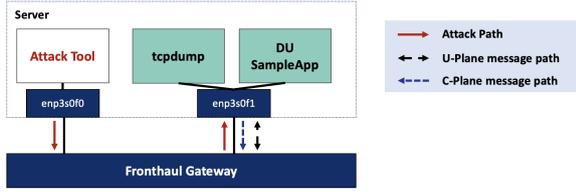

Fig. 2. Traffic test architecture

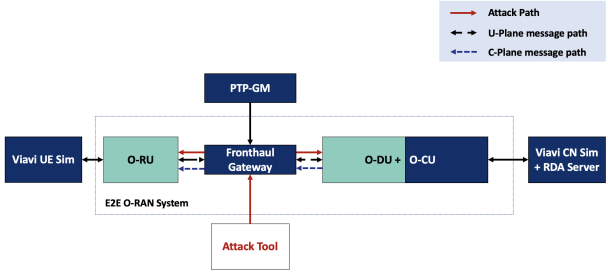

Fig. 3. Topology 1 consisting of Viavi UE simulator [19], vendor 1 O-RU, commercial fronthaul gateway, vendor 1 O-DU, vendor 1 O-CU, Viavi CN simulator, and RDA server

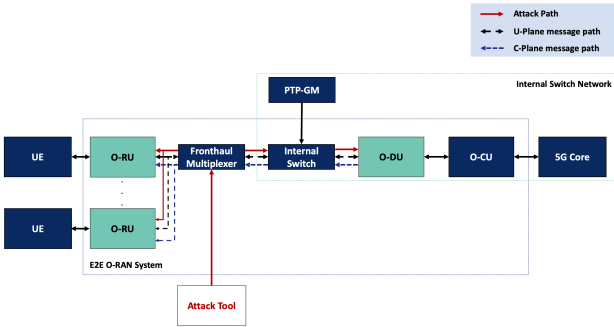

Fig. 4. Topology 2 with real UE, vendor 2 O-RU, commercial fronthaul multiplexer, commercial fronthaul switch, vendor 2 O-DU, vendor 2 O-DU, and commercial 5G Core

## IV. EXPERIMENTAL RESULTS

Experiments were conducted to verify the following:
- Whether our developed C/U-Plane attack tool meets the requirements specified in [5], in particular, whether it can well support new features we claimed in Section II and Section III.
- Evaluate the impacts of DoS attacks targeting C/U-Plane of both O-DU and O-RU in end-to-end systems using commercial O-RAN products.

Figure 2 is a traffic test architecture that will be used to test our C/U-plane attack tool functionality before it is used in an end-to-end system. Figure 3 and Figure 4 show two different topologies used in the experiments for achieving our second evaluation objective. Figure 3 is used as the main topology where most of the data is collected from, while Figure 4 is only used to show how different systems can have different protection levels based on the design by developers. In topology 1, we have a commercial O-RAN system with Viavi TM500 for user equipment (UE) simulation and Viavi TeraVM for core emulation. O-RU, O-DU, and precision time protocol grandmaster (PTP-GM) are connected through a fronthaul gateway. In topology 2, we have a commercial O-RAN system with a commercial 5G core and real UE. O-DU is connected to seven O-RUs through a fronthaul multiplexer and an internal switch, where each O-RU have UE connected to them. Due to the confidentiality agreement, we are unable to provide more details regarding the components involved in both topologies. In both environments, we run the attacker for 30 seconds since, in most cases, when the attack causes degradation, the impact will be apparent in less than that duration. Note that there's a limitation where the experiment is only conducted once in this research.

Two scenarios were investigated. Scenario I was designed to verify the effectiveness of Contribution 1. Scenario II was designed to evaluate the impact of Contribution II. We define the following scenarios:

Scenario I:
1) Verify our tool's compliance to section 7.2.2 on supporting three variations of MAC addresses (spoofed MAC of O-RU(s), random MACs, and broadcast)
2) Verify our tool's compliance to section 7.2.2 on supporting volumetric tiers (10Mbps, 100Mbps, 1Gbps)
3) Verify our tool's capability to send U-Plane packets

Scenario II:
1) TIFG 7.2.2 C-Plane eCPRI DoS Attack (Sub-scenario 1): In this sub-scenario will compare the impact of TIFG 7.2.2 C-Plane eCPRI DoS attack towards O-DU in two different environments, topology 1 (Figure 3) and topology 2 (Figure 4).
- Experiment: the attack tool generates six combinations of attack using eCPRI real-time control data message over Ethernet based on the test procedure in the specification, where there are two DoS source address variations (spoofed MAC of O-RU(s), random source MACs) and three volumetric tiers (10Mbps, 100Mbps, 1Gbps).
- Exploited DUT(s): O-RU and O-DU
- Observation: the impact will be measured based on whether the system is down or not through monitoring the system. In environment A, we can observe this based on the throughput degradation and log in O-DU. In environment B, the system monitor shows whether the system is up or down, as well as throughput when the system is up.

2) Attack towards O-DU and O-RU (Sub-scenario 2): In this sub-scenario, O-RU and O-DU in topology 1 (Figure 3) will exchange C/U-plane packet normally. The attack will launch different combinations of attacks based on different message types, source MAC addresses, and volume to find possible loopholes in the packet processing of DUT(s).

Fig. 5. C/U-Plane DoS attack can generate varying traffic types

- Experiment: the attack tool generates message type(s) targeting packet processing flaws of O-DU and O-RU. The message type tested in our experiment is C-Plane downlink, U-Plane downlink, and U-Plane uplink. The messages/packets will be sent using two DoS source address variations (spoofed MAC of O-RU(s), random source MACs) and three volumetric tiers (10Mbps, 100Mbps, 1Gbps).
- Exploited DUT(s): O-DU, O-RU.
- Observation: the impact will be measured in terms of aggregate downlink and uplink shared channel throughput. In addition, we observe the behavior from DUT through the log, including PHY to MAC UL BLER.

### A. DoS Traffic Types, Source Address, and Volumetric Tiers

We set up the environment in Figure 2 to validate the variation of traffic types, source addresses, and volumetric tiers. In validating traffic types and source addresses, we use tcpdump [20] to capture traffic sent by the C/U-Plane attack tool. The C/U-plane attack tool is linked to enp3s0f0 and runs to send each traffic type. In this test, we consider the source MAC address to be 00:11:22:33:44:55 and the destination MAC address to be 00:11:22:33:44:66. We captured the traffic of running our attack tool by using tcpdump on enp3s0f1. Figure 5 shows the captured traffic. The upper part is eCPRI real-time control data message which, in this case, we use a C-Plane downlink message, the middle part is a U-Plane downlink, and the bottom part is a U-Plane uplink.

We use the same method and addressing consideration to capture the source address variation by running the C/U-Plane attack tool linked to enp3s0f0 and tcpdump on enp3s0f1. Figure 6 shows the captured traffic. The upper part shows the spoofed MAC address of O-RU (in this case, it is the assumed source address), the middle part is random MACs, and the bottom part is broadcast MAC.

Fig. 6. C/U-Plane DoS attack can generate varying source MAC addresses

Fig. 7. C/U-Plane DoS attack can generate varying volumetric tiers

In proving that the C/U-Plane attack tool can send the intended volumetric tiers to DUT(s), we use DU SampleApp. DU SampleApp is a part of RU-DU emulators provided by the OSC O-DU Low project [15]. DU SampleApp can calculate the received traffic that can be observed from the log. In our test, we didn't connect RU SampleApp to DU SampleApp in order to provide the exact kbps received from the C/U-Plane attack tool since if we connect to RU SampleApp, the resulting log will be mixed with RU SampleApp traffic. Figure 7 shows the log of DU SampleApp under attack. The upper part is when we use the C-Plane message to attack DU SampleApp using 10 Mbps, 100 Mbps, and 1 Gbps, respectively. We can see the approximate traffic received by DU SampleApp is 9.896 Mbps, 98.966 Mbps, and 989.664 Mbps. Note that it is the received traffic counted by DU SampleApp; there might be overhead along the path that causes the reduction of speed. The lower part is when we use the U-Plane message to attack with a retrieval rate of 9.830 Mbps, 98.304 Mbps, and 983.040 Mbps, respectively.

### B. C-Plane DoS Attack (TIFG 7.2.2) in Two Settings

We compared two different commercial settings: topology 1 (Figure 3) and topology 2 (Figure 4). We run our attack tool for 30 seconds to send eCPRI real-time ctrl data message

| Destination MAC | | | O-DU | | | | | | | | |
|---|---|---|---|---|---|---|---|---|---|---|---|
| Source MAC | | | O-RU | | | Random MACs | | | O-DU | | |
| Volume | | | 10 Mbps | 100 Mbps | 1000 Mbps | 10 Mbps | 100 Mbps | 1000 Mbps | 10 Mbps | 100 Mbps | 1000 Mbps |
| Traffic Type | C-Plane | Downlink | PASS | PASS | PASS | PASS | PASS | PASS | FAIL | FAIL | FAIL |
| | U-Plane | Downlink | PASS | PASS | PASS | FAIL | FAIL | FAIL | PASS | FAIL | FAIL |
| | | Uplink | PASS | PASS | PASS | FAIL | FAIL | FAIL | FAIL | FAIL | FAIL |

| Destination MAC | | | O-RU | | | | | | | | |
|---|---|---|---|---|---|---|---|---|---|---|---|
| Source MAC | | | O-DU | | | Random MACs | | | O-RU | | |
| Volume | | | 10 Mbps | 100 Mbps | 1000 Mbps | 10 Mbps | 100 Mbps | 1000 Mbps | 10 Mbps | 100 Mbps | 1000 Mbps |
| Traffic Type | C-Plane | Downlink | FAIL | FAIL | FAIL | FAIL | FAIL | FAIL | PASS | PASS | PASS |
| | U-Plane | Downlink | FAIL | FAIL | FAIL | PASS | FAIL | FAIL | PASS | PASS | PASS |
| | | Uplink | FAIL | FAIL | FAIL | PASS | PASS | PASS | PASS | PASS | PASS |

**Severity Levels**

High
- O-DU Low crashed
- Immediately broke the E2E connection (UE lost connection)
- +- 10 seconds before it broke the E2E connection (UE lost connection)
- Severe degradation of UL throughput and disruption on DL throughput
- Minimal disruption on UL and DL throughput that totally recovers once attack stops

Low

Fig. 8. Experimental result: impact of attack towards O-DU and O-RU

TABLE I
C-PLANE DOS ATTACK (TIFG 7.2.2) IN TWO DIFFERENT
ENVIRONMENTS USING O-RU'S MAC AS SOURCE MAC ADDRESS

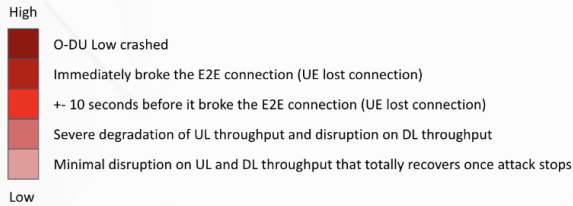

| Source MAC | O-RU | | |
|---|---|---|---|
| | 10 Mbps | 100 Mbps | 1 Gbps |
| Topology 1 | PASS | PASS | PASS |
| Topology 2 | FAIL | FAIL | FAIL |
| Source MAC | Random MACs | | |
| | 10 Mbps | 100 Mbps | 1 Gbps |
| Topology 1 | PASS | PASS | PASS |
| Topology 2 | FAIL | FAIL | FAIL |

over Ethernet (C-Plane message) to O-DU as the Device Under Test (DUT). Referencing to the newest specification [5], the upper part of Table I shows the result when using O-RU MAC address as source MAC and the lower part of Table I shows the result when using random source MACs.

Topology 1 demonstrated its successful compliance with the protection requirement for the C-Plane, effectively thwarting the malicious traffic from our attack tool without any adverse impact on traffic performance or system stability. However, in contrast, topology 2 failed to withstand this test case, leading to a system outage.

### C. Impact of C/U-Plane DoS Attack Towards O-DU

We conduct further attacks on C/U-Plane in topology 1 by using different message types, source MAC addresses, and volumes towards O-DU. Figure 8 shows the PASS or FAIL result of the attack conducted to O-DU's C/U-Plane VF targeting C-Plane packet processing and U-Plane packet processing, respectively. 'PASS' means no degradation is observed in the end-to-end system, while 'FAIL' means degradation is observed in the end-to-end system. In Figure 8, we also show the severity level of attack impact based on the result of observation. We observed the following anomalies:

1) Attack using C-Plane as traffic type only failed the system when O-DU's MAC address is used as source MAC address.
2) Attack using U-Plane as traffic type failed the system when random MACs and O-DU's MAC are used as source MAC address.

We conducted tests to observe the system's reaction to U-Plane DL and U-Plane UL messages as the DoS traffic types, using random MAC addresses as the source MAC addresses. Figure 9 illustrates the aggregate downlink and uplink throughput (in kbps) for the 'FAIL' cases. In the 10 Mbps scenario, there were slight delays in throughput drop. Meanwhile, in 100 Mbps and 1 Gbps cases, the throughput dropped almost immediately.

In our observations, we noted the recurrence of the same error log (Figure 10) in O-DU for all volumetric U-Plane traffic types (downlink and uplink), leading to its automatic restart. Interestingly, O-DU continued to process the DoS packets, even when they were sent with random MAC addresses as the source MAC addresses. This observation highlights that in the system, O-DU does not take into account or differentiate

Fig. 9. Impact of U-Plane DoS attack with spoofed random MAC address as source address

Fig. 11. Impact of DoS attack with spoofed O-DU MAC address as source MAC address

Fig. 10. Impact of U-Plane Vulnerabilities to O-DU (log)

Fig. 12. UL BLER on O-DU

the source MAC address and continues to process the packet regardless. In the next case, we examine the scenario where the O-DU's C/U-Plane is targeted using the O-DU's MAC address. In topology 1, the O-RU port is configured as static, while both the O-DU and the attack tool operate in MAC learning mode. According to this concept, the fronthaul gateway determines the MAC address of a port based on the source MAC address of outgoing packets from that port. Therefore, when the attack tool assumes the O-DU's MAC address and both the O-DU and the attack tool are sending packets, the fronthaul gateway associates the O-DU's MAC address with the latest port that it receives a packet that carries O-DU's MAC address as the source address — i.e., we are launching a MAC spoofing attack [21] [22].

Thus, wrong forwarding will happen in the switch when O-RU sends packets to O-DU because the fronthaul gateway cannot identify which one is the real O-DU. This caused the drop of aggregate downlink/uplink throughput (Figure 11), rising uplink block rate error (UL BLER) of PHY-to-MAC calculated by O-DU (Figure 12), and in most of 1 Gbps case, UL BLER reached to 100%, resulting UE losing connection (Figure 13). We also observed a point where using low volume (10 Mbps) when sending U-Plane downlink using O-DU's MAC address as the source MAC address didn't affect the system, while raising it to 100 Mbps affected it.

### D. Impact of C/U-Plane DoS Attack Towards O-RU

We conduct further attacks on C/U-Plane in topology 1 by using different message types, source MAC addresses, and volumes towards O-RU. Some of anomalies we see from Figure 8:

Fig. 13. Record on O-DU

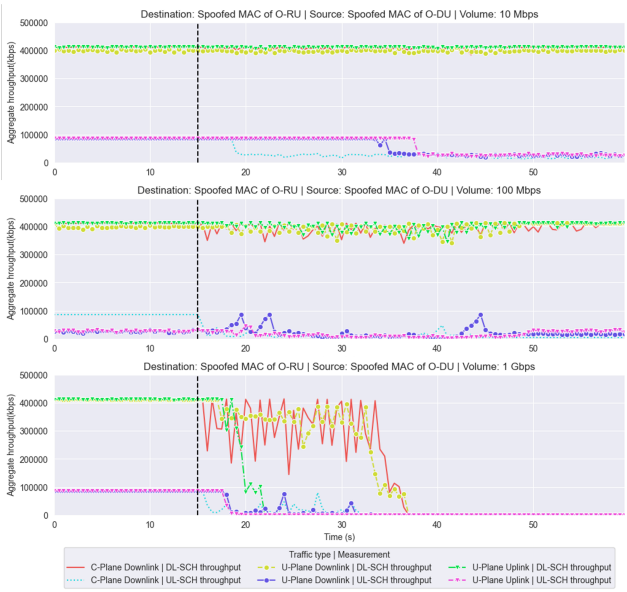

Fig. 14. Impact of O-RU Attack using O-DU's MAC

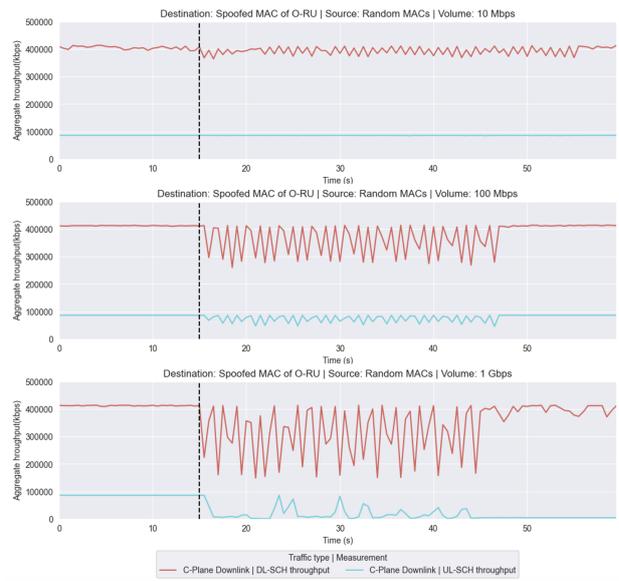

Fig. 15. Impact of O-RU attack using C-Plane and random MACs

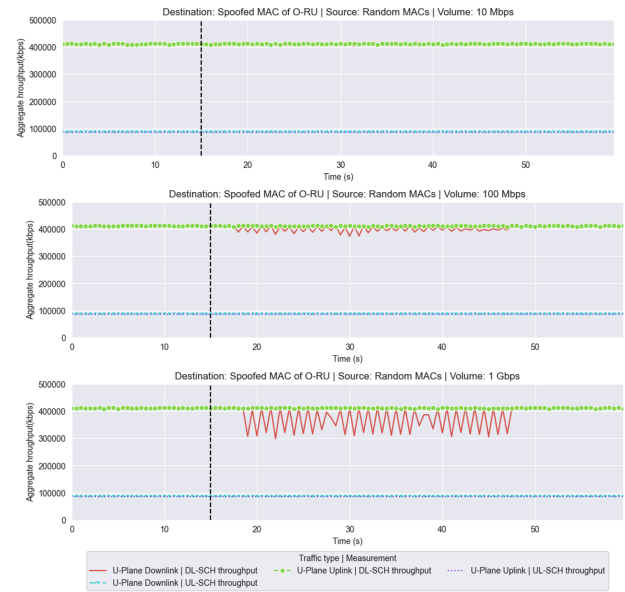

Fig. 16. Impact of O-RU attack using U-Plane and random MACs

1) It failed when the packet was sent using O-DU's MAC address as the source MAC address, which includes the C-Plane downlink, and U-Plane downlink / uplink.
2) It failed when C-Plane traffic is sent using random source MACs in all volumetric tiers.
3) It failed when U-Plane downlink traffic is sent using random source MACs in volume 100 Mbps and 1 Gbps.

On C/U-Plane attack with spoofed O-DU's MAC (Figure 14), it appears that lower rate of attack (like 10 Mbps) mainly impacts the uplink throughput (as shown in UL-SCH) and for the higher rate of attack (i.e., 1 Gbps), both the uplink and downlink throughput is impacted. Furthermore, for both uplink and downlink, under higher rate of attacks, the system does not recover even when the attack stopped.

On C-Plane with random MACs (Figure 15), we observe that in 10 Mbps and 100 Mbps, the system recovered after the attack was stopped. Meanwhile, the aggregate downlink throughput is affected significantly when 1 Gbps is sent, along with non-recoverable aggregate UL-SCH throughput. On the U-Plane side (Figure 16), there's minimal disruption to the system, only caused by U-Plane Downlink, and it only caused a slight disturbance on DL-SCH throughput, even at 1 Gbps. The DL-SCH throughput also recovered after the attack was stopped.

The biggest impact is seen when O-DU's MAC address is used (Figure 14), where it most likely happens due to the switch table is affected by our attack traffic and causes the switch to forward the traffic for O-DU to the host that runs our attack tool.

## V. Conclusion and Future Works

In this paper, we have explained how we developed a C/U-Plane attack tool to observe the impact of DoS attacks against C/U-Plane of O-DU and O-RU. Our tool is fully compliant with O-RAN E2E test specification [5] and able to be expanded to support varied test cases. Using our tool, we launched different attack scenarios in an end-to-end environment using different traffic types, rates, and source MAC addresses. We showed the different security levels of two DUTs in their behavior against specification-based DoS Attacks. We also conducted various combinations of attack. We found several issues in both O-DU's and O-RU's packet processing. Some result in insignificant degradation, but some can crash the targeted system.

However, this research currently only showed the results of attacks using three kinds of messages with the variation of MAC address and volume per second,. Thus, the targeted impact on DUT(s) is still limited to the processing of the source MAC address, packet type, and packet direction since the other fields are of random values. For future work, we plan to conduct more systematic tests to investigate other possible vulnerabilities in the open fronthaul implementation.

## Acknowledgement

This research is supported in part by the National Science and Technology Council, Taiwan, under grant number 111-2221-E-011-069-MY3, 112-2218-E-011-004, and 112-2218-E-011-006, and in part by the National Research Foundation, Singapore and Infocomm Media Development Authority under its Future Communications Research & Development Programme (FCP-SUTD-RG-2021-016). Any opinions, findings, and conclusions or recommendations expressed in this material are those of the author(s) and do not reflect the views of the National Science and Technology Council, Taiwan, and the National Research Foundation and Infocomm Media Development Authority, Singapore.